\documentclass[
aps,%
10pt,%
final,%
notitlepage,%
oneside,
twocolumn,%
nobibnotes,%
nofootinbib,%
superscriptaddress,%
noshowpacs,%
centertags]%
{revtex4}

\begin{document}
\selectlanguage{english}


\title{Molecular Emission in Dense Massive Clumps from the Star-Forming Regions S231-S235}

\author{\firstname{D.~A.}~\surname{Ladeyschikov}}
% \email{dmitry.ladeyschikov@urfu.ru}
 \affiliation{Kourovka Astronomical Observatory, Ural Federal University, Ekaterinburg, 620000 Russia}

 \author{\firstname{M.~S.}~\surname{Kirsanova}}
% \email{kirsanova@inasan.ru}
 \affiliation{Institute of Astronomy of the Russian Academy of Sciences, Moscow, 119017 Russia}

\author{\firstname{A.~P.}~\surname{Tsivilev}}
% \email{tsivilev@prao.ru}
 \affiliation{Pushchino Radio Astronomy Observatory, Lebedev Physical Institute of the Russian Academy of Sciences, Pushchino, 142290 Russia}

\author{\firstname{A.~M.}~\surname{Sobolev}}
% \email{andrej.sobolev@urfu.ru}
 \affiliation{Kourovka Astronomical Observatory, Ural Federal University, Ekaterinburg, 620000 Russia}

%\received{October 9, 2015}  \revised{March 21, 2016}

\begin{abstract}
The article deals with observations of star-forming regions S231-S235
in 'quasi-thermal' lines of ammonia (NH$_3$), cyanoacetylene (HC$_3$N) and maser lines of
methanol (CH$_3$OH) and water vapor (H$_2$O). S231-S235 regions is
situated in the giant molecular cloud G174+2.5. We selected all massive
molecular clumps in G174+2.5 using archive CO data. For the each clump
we determined mass, size and CO column density. After that we performed
observations of these clumps. We report about first detections of
NH$_3$ and HC$_3$N lines toward the molecular clumps WB89\,673 and
WB89\,668. This means that high-density gas is present there. Physical
parameters of molecular gas in the clumps were estimated using the data
on ammonia emission. We found that the gas temperature and the hydrogen number density
are in the ranges 16-30~K and 2.8-7.2$\times10^3$~cm$^{-3}$,
respectively. The shock-tracing line of CH$_3$OH molecule at 36.2~GHz is
newly detected toward WB89\,673. 
%The paper is concerned with the study of the star-forming regions S231--S235 in radio lines of molecules of the interstellar medium---carbon monoxide (CO), ammonia (NH$_3$), cyanoacetylene (HC$_3$N), in maser lines---methanol (CH$_3$OH) and water vapor (H$_2$O). The regions S231--S235 belong to the giant molecular cloud G174+2.5. The goal of this paper is to search for new sources of emission toward molecular clumps and to estimate their physical parameters from CO and NH$_3$ molecular lines. We obtained new detections of NH$_3$ and HC$_3$N lines in the sources WB89\,673 and WB89\,668 which indicates the presence of high-density gas. From the CO line, we derived sizes, column densities, and masses of molecular clumps. From the NH$_3$ line, we derived gas kinetic temperatures and number densities in molecular clumps. We determined that kinetic temperatures and number densities of molecular gas are within the limits $16$--$30$~K and $2.8$--$7.2\times10^3$ cm$^{-3}$ respectively. The shock-tracing line of CH$_3$OH molecule at a frequency of 36.2 GHz was detected in WB89\,673 for the first time.
\end{abstract}

\maketitle

\section{INTRODUCTION}

One of the most topical and rapidly developing fields in astrophysics is the study of star-forming regions. A great number of molecules in the interstellar medium which are quite intensively radiating in radio lines gives a plenty of possibilities for this. It is considered currently that star formation occurs in regions of high concentration of molecular gas, clumps, the main component of which is molecular hydrogen~(H$_2$). As hydrogen molecules in clumps do not radiate in the radio range, radio lines of other molecules are used to trace the presence of molecular gas, ongoing processes, and conditions in the interstellar medium. Particularly, molecular lines of carbon monoxide (CO) show general distribution of molecular gas in the star-forming regions of our Galaxy~\cite{Combes91}. Lines of ammonia (NH$_3$) molecule are tracers of temperature~\cite{Walmsley_1983} and gas high density~\cite{Jijina99}. Molecular lines of cyanoacetylene (HC$_3$N) are also gas high density tracers~\cite{Morris76}. The observed data shows that maser and ``quasi-thermal'' lines of methanol (CH$_3$OH) make it possible to investigate outflows from young stellar objects and shock waves in the interstellar medium~\mbox{\cite{Sobolev07,Salii06,Sutton04,Salii02,Voronkov06}}, and masers at the transition of water molecule (H$_2$O) indicate ongoing active star-forming
processes~\cite{Gray12}. On the whole, the present information provides an opportunity of a comprehensive study of star-forming regions including estimation of their physical parameters as well.
 
The goal of this paper is to study massive molecular clumps of the star-forming regions S231--S235 situated in the giant molecular cloud (GMC)  G174+2.5. In this direction, there are four advanced regions of ionized hydrogen: S231, S232, S233, and S235 in accordance with the catalog by Sharpless~\cite{Sharpless59}. Researchers distinguish six well-studied young star clusters: S235\,Central, S235\,East\,1, S235\,East\,2, and S235\,AB (see, e.g.,~\cite{Kirsanova08,Kirsanova14,Chavarria14}),
S233\,IR (see~\cite{Ginsburg09,Leurini07,Beuther07}),
and G173.57+2.43 (see~\cite{Chakraborty00,Shepherd02}). Cross-identification of clusters called differently by various authors is given in the paper by Camargo~\cite{Camargo11}. On the periphery, there are the less-studied star-forming regions WB89\,673 and WB89\,668 which are named from the Wouterloot--Brand catalog~\cite{Wouterloot89}. Previously Heyer et al.~\cite{Heyer96} studied the morphology and kinematics of this star-forming complex from the CO radio lines. The maps of CO line emission show that the most prominent star-forming regions in the cloud G174+2.5 are S231--S235. Distance estimates to them range from $1.5$ to $2.3$~kpc~\cite{Burns15,Reipurth08}. The regions S231--S235 are situated in the direction close to the galactic anticenter. As follows from the paper by Dame et al.~\cite{Dame_2001}, no such star-forming regions more distant from the Sun are observed in this direction in the Galaxy.

In the present paper, we present the results of observations of molecular radio lines in molecular clumps, previously selected using the archive data on the CO line emission in the giant molecular cloud G174+2.5, which have been carried out at the radio telescope RT-22 of the Puschino Radio Astronomy Observatory of the Lebedev Physical Institute of the Russian Academy of Sciences (PRAO ASC LPI). First, we searched for class I methanol masers in the direction to the selected molecular clumps. After detection of them, to prove the presence of dense gas and to determine physical parameters of gas in the clumps, density and temperature first of all, we conducted observations in the NH$_3$ and HC$_3$N lines. Thus, in this paper we present general characteristics of emission of molecules toward the star-forming regions S231--S235 and estimations of the physical parameters of the gas and of the mass of molecular clumps. Instruments, sources, and methods we used are described in the corresponding Sections. 

\section{SELECTION OF THE OBJECTS FOR OBSERVATION}\label{sec:selection}
\begin{figure*}
   \setcaptionmargin{5mm} \onelinecaptionsfalse \captionstyle{normal}
  \center
        \includegraphics [scale=0.87] {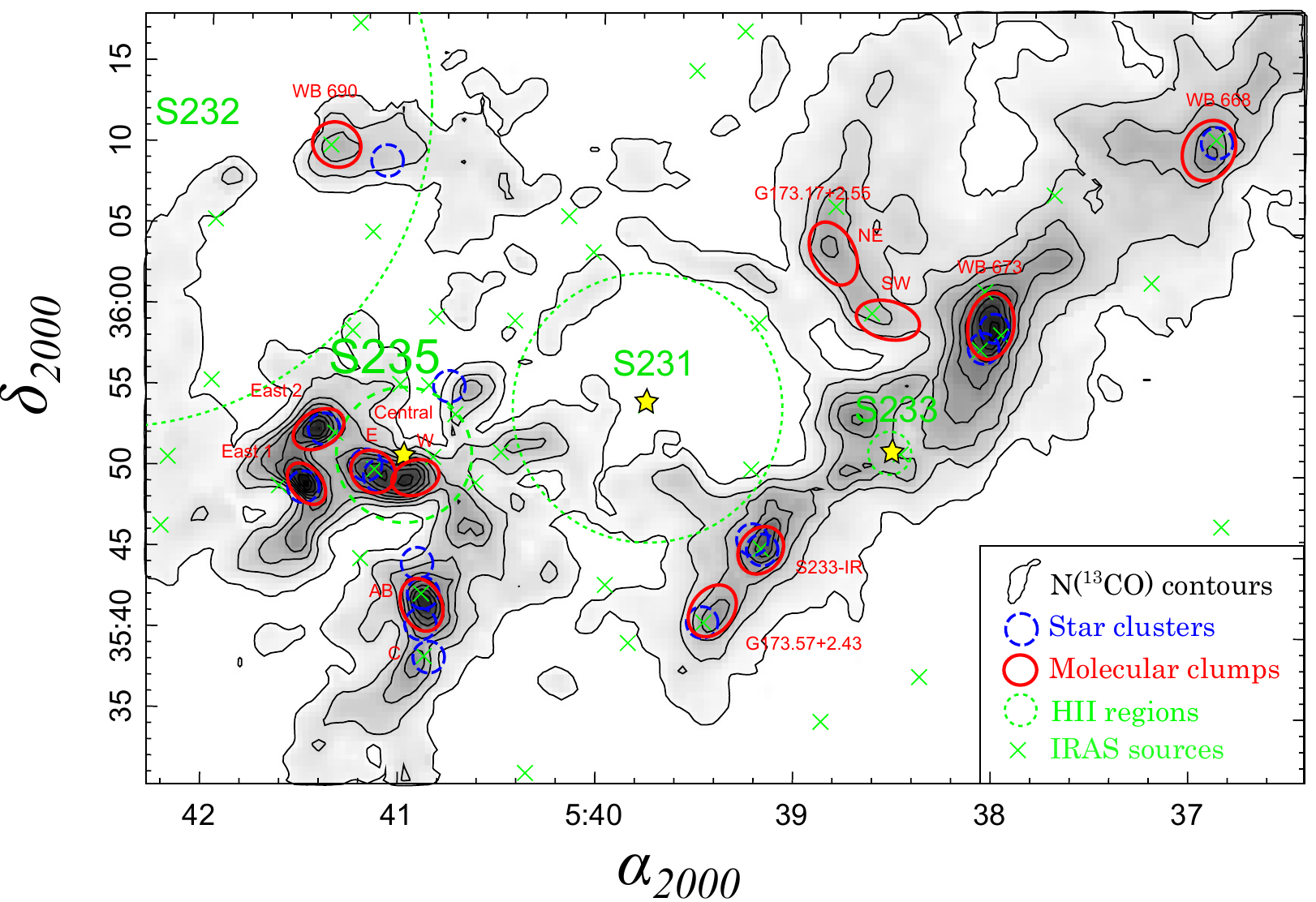}
 \caption[Distribution of column density of $^{13}$CO molecule toward the regions S231--S235]{Distribution of column density of $^{13}$CO toward the star-forming region S231--S235. Outer contour corresponds to the value $5\times10^{15}$~cm$^{-2}$, the most inner---$6\times10^{16}$~cm$^{-2}$, contour step is $8.4\times10^{15}$~cm$^{-2}$. Crosses denote the positions of the IRAS sources. Red ellipses mark molecular clumps selected within the frame of the present paper for observations in the lines of methanol, cyanoacetylene, and ammonia. The sizes of ellipses correspond to the sizes of molecular clumps. Blue dashed circles denote the positions of young star clusters from the data by Camargo et al.~\cite{Camargo11}. Green dashed circles mark the regions of ionized hydrogen (H\,II). The radii of the circles correspond to the radii of the H\,II{} regions from the DSS R images. Star-shaped figures show the position of stars ionizing the H\,II regions.}
\label{img:CO}
\end{figure*}

We used the archive data with observations in the $^{12}$CO\mbox{(1--0)} and $^{13}$CO\mbox{(1--0)} lines in order to select molecular clumps toward the star-forming regions \mbox{S231--S235} to estimate their physical characteristics. They are described in Section~\ref{sec:data_CO}. The details of method of physical parameters' calculation are given in Annex~\ref{app:co}. We overlaid the positions of the IRAS sources onto the derived map of column density of $^{13}$CO toward the star-forming regions \mbox{S231--S235} and limited the research area with a circle of a radius of 50\arcmin{} around the center of ionized hydrogen S231 (\mbox{$\alpha_{\rm J2000} = 5^{\rm h} 39^{\rm m} 45^{\rm s}$}, \mbox{$\delta_{\rm J2000}=35\degr 54\arcmin 02\arcsec$}) %\ahmsB{5}{39}{45}  \ddmsB{35}{54}{02}
(see Fig.~\ref{img:CO}).

For observations in the NH$_3$  line, we selected the positions of centers of the~$^{13}$CO column density peaks. In the CH$_3$OH and HC$_3$N lines, the sources were observed at the positions which coincide or are quite close  to local peaks of $^{13}$CO column density. In the case of close proximity of several IRAS sources to the peak, if the angular distance between them was smaller than the beam size of the RT-22, then we chose the source closest to the $^{13}$CO peak as a pointing source. As a result of this analysis, we visually selected 10 molecular clumps to observe at the RT-22. We added a calibration source  Dr21 to our sample for comparing the observed results with other works.

Table~\ref{tbl:sources} shows the coordinates and general physical characteristics of the sources. Radial velocity and clump sizes are derived from the data on emssion in the $^{13}$CO(1--0) line. Excitation temperature is calculated using the data on emssion in the $^{12}$CO(1--0) line, its average value is shown within the limits of molecular clump sizes. Numerical values of column density are calculated on the basis of the map of its distribution presented in Fig~\ref{img:CO}. The table shows average column densities of H$_2$ and masses within the limits of molecular clump sizes. The last column gives virial parameter $\alpha_{\rm vir} \equiv M_{\rm vir}/M$ (see the description in Section~\ref{sec:mass}).

 %Table 1
\begin{table*}
 \setcaptionmargin{0mm} \onelinecaptionsfalse \captionstyle{normal}
\caption[List of sources to observe.]{Catalog of molecular clumps in the $^{13}$CO line with general physical parameters (see the text for details).
} \label{tbl:sources}
\medskip
% \footnotesize
\begin{tabular}{l|c|c|c|c|c|c|c|c|c|c|c} \hline
~~~~~~~Name & $\alpha_{J2000}$, & $\delta_{J2000}$, &
$\theta_{\rm FWHM}$, & $R$, &  $V_{\rm LSR}$, &  $\Delta V$, &
$T_{\rm ex}$, & $\bar{N}({\rm H}_2)$, &  $M({\rm H}_2)$,
% & $n({\rm H}_2)$
& $\tau(^{13}{\rm CO})$
& $\alpha_{\rm vir}$  \\
%    & \ahms{}{~}{~}  & \ddms{}{~}{~} & \arcmin & �� & ��\,�$^{-1}$  &  ��\,�$^{-1}$  & K &   $10^{22}$ ��$^{-2}$ & ${\rm M}_{\odot}$   &     &    \\ \hline
& ${\rm h}$ ${\rm m}$ ${\rm s}$ & $\degr$ $\arcmin$ $\arcsec$ &
$\arcmin$ & pc & km\,s$^{-1}$ & km\,s$^{-1}$  & K &   $10^{22}$
cm$^{-2}$ & ${\rm M}_{\odot}$   & &
\\ \hline
WB89\,690       & 5~41~21.6 & +36~10~00 & 3.1$\times$2.7 & 0.87 &  $-$21.0  & 1.83   & 16.5  & 1.45 & 733   & 0.65  & 0.86  \\  %
WB89\,668       & 5~36~54.3 & +36~10~16 & 3.9$\times$3.1 & 1.06 &  $-$17.2  &  2.70  & 14.1  & 1.60 & 1199  & 0.85  & 1.40\\  %
WB89\,673       & 5~38~00.6 & +35~59~17 & 4.1$\times$2.8 & 1.04 & $-$19.5   & 3.16   & 20.8 & 2.90 & 2112   & 0.74 & 1.07 \\ %
G173.17+2.55 NE & 5~38~49.0 & +36~03~41 & 4.1$\times$2.6 & 0.99 & $-$17.9 &  2.44  & 16.4  & 1.66 & 1095    & 0.69 & 1.18 \\ %
G173.57+2.43    & 5~39~24.7 & +35~41~28 & 3.5$\times$2.6 & 0.90 &  $-$16.8 &  2.38  & 17.2  & 1.70 & 932   & 0.65 & 1.20 \\ %
S233-IR         & 5~39~10.2 & +35~45~15 & 3.1$\times$2.6 & 0.83 & $-$16.9  &   2.76  & 22.1  & 2.29 & 1048   & 0.55 & 1.31 \\ %
S235\,Central E & 5~41~08.8 & +35~49~47 & 2.9$\times$2.5 & 0.83 & $-$19.6 &  1.92 & 35.6 & 3.67 & 1683    & 0.35 & 0.39 \\
S235\,Central W & 5~40~55.8 & +35~49~27 & 3.0$\times$2.1 & 0.74 & $-$21.5 & 1.93 & 31.6 & 3.65 & 1347   & 0.49 & 0.45 \\
S235\,East1     & 5~41~29.0 & +35~48~58 & 2.9$\times$1.9 & 0.74 & $-$18.9  &   1.80 & 32.7  & 4.31 & 1591   & 0.51 & 0.33  \\ %
S235\,East2     & 5~41~25.6 & +35~52~21 & 3.4$\times$2.2 & 0.87 & $-$20.8 &    1.78 & 29.4  & 3.62 & 1822   & 0.55 & 0.33 \\ %
S235-AB         & 5~40~53.3 & +35~41~35 & 3.4$\times$2.5 & 0.90 & $-$16.5  &    2.30 & 27.0  & 3.53 & 1935   & 0.59 & 0.54 \\ %
\hline
  \end{tabular}
%   \normalsize
\end{table*}

All the selected molecular clumps radiate in the continuum at a wavelength of 1.12~mm from the Bolocam~\cite{Ginsburg13} survey data which is indicative of heated dense gas in them. Almost all of them correspond to young star clusters according to the data from infrared sky surveys Wide-Field Infrared Survey Explorer (WISE,~\cite{Wright10}) and UKIRT Infrared Deep Sky Survey (UKIDSS,~\cite{Lawrence07}). Section~\ref{sec:star.formation} in detail considers the association of clumps and clusters.

\section{ARCHIVE DATA ON THE CO EMISSION} \label{sec:data_CO}

Data on emssion in the $^{12}$CO\mbox{(1--0)} and $^{13}$CO(1--0) lines are obtained using the results of observations within the program of highly-accurate survey of the Galactic plane in the line of the CO molecule~\cite{Mottram15} conducted at the 13.7-m telescope of the Five College Radio Astronomy Observatory (FCRAO) with the Second Quabbin Observatory Imaging Array (SEQUOIA), a 32-pixel focal receiver. The regions \mbox{S231--S235} were mapped in January of 2000. The map of line emssion of both CO isotopes covers the region with a size of 150\arcmin$\times$150\arcmin\ with the center at \mbox{$l=173\fdg25$}, \mbox{$b=2\fdg75$} (\mbox{$\alpha_{\rm J2000} = 5^{\rm h} 40^{\rm m}$} \mbox{$\delta_{\rm J2000}=36\degr 07\arcmin$}). The beam size for this telescope is 45\arcsec\ for \mbox{$^{12}$CO(1--0)} and 47\arcsec\ for \mbox{$^{13}$CO(1--0)}. Observed frequency was set 115.27120~GHz for the \mbox{$^{12}$CO(1--0)} line and 110.20135~GHz for \mbox{$^{13}$CO(1--0)}. Spatial step was 22\farcs5\ which is two times smaller than the half-power beamwidth (HPBW). Velocity step was equal to 0.127~km\,s$^{-1}$ for $^{12}$CO(1--0) and 0.133~km\,s$^{-1}$ for $^{13}$CO(1--0). The noise level $\sigma_{T_{\rm mb}}$ for $^{12}$CO(1--0) was 1.1~K and for $^{13}$CO(1--0)---0.63~K on  main-beam temperature scale. For the reduction of data in the CO lines and the calculation of physical parameters, we used the {\tt MIRIAD} package~\cite{Sault95}. Integrating and statistical analysis of physical parameters were carried out with the package {\tt ds9}~\cite{Joye03}.

\section{OBSERVED DATA}\label{sec:obs}

\begin{figure*}%[p]
   \setcaptionmargin{5mm} \onelinecaptionsfalse \captionstyle{normal}
        \includegraphics [scale=0.8] {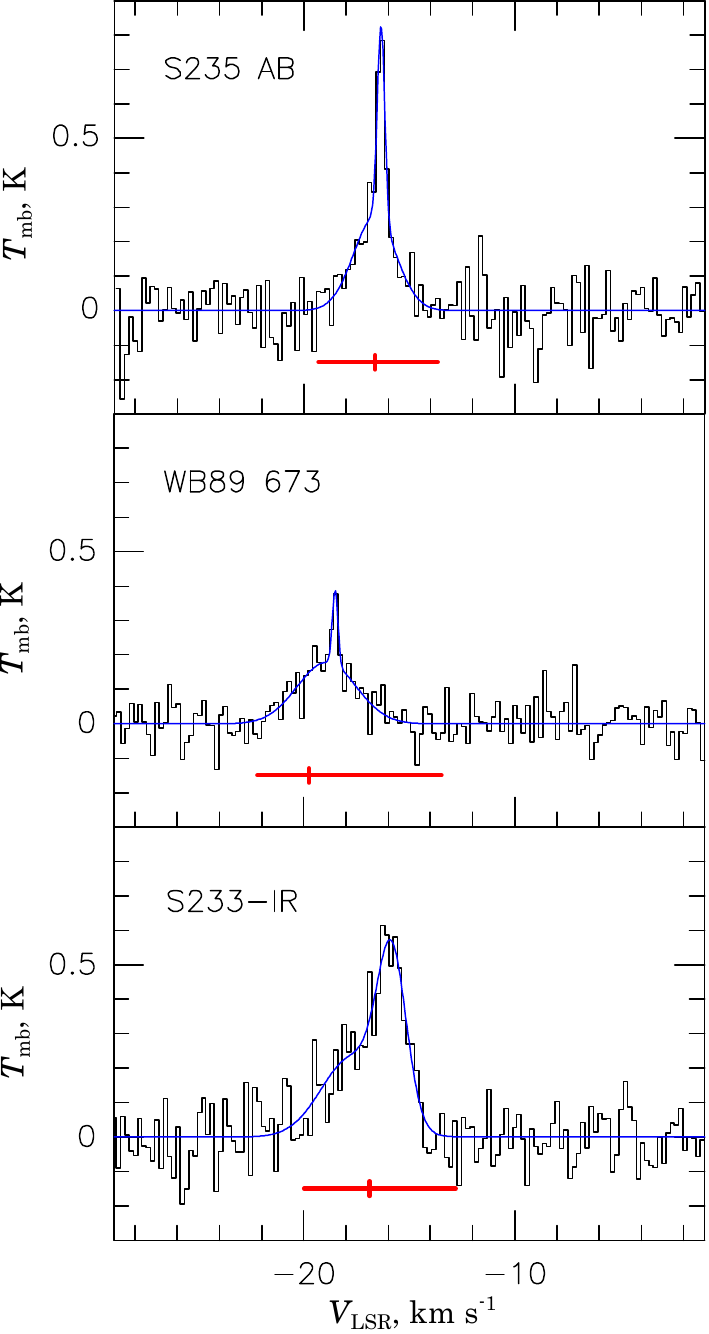}
  \caption[Source spectra in the CH$_3$OH line.]{Spectra of the detected sources in the CH$_3$OH line at a frequency of 36.2 GHz. A blue line denotes approximation of line profiles with the Gaussian function, a red line segment at the lower part of spectra shows the velocity range of $^{13}$CO, and the peak of the $^{13}$CO line is marked with a tick.}

  \label{img:ch3oh}
\end{figure*}

\begin{figure*}%[p]
 \setcaptionmargin{5mm} \onelinecaptionstrue \captionstyle{normal}
  \center
  \begin{tabular}{@{}cccc@{}}
        \includegraphics [scale=0.8] {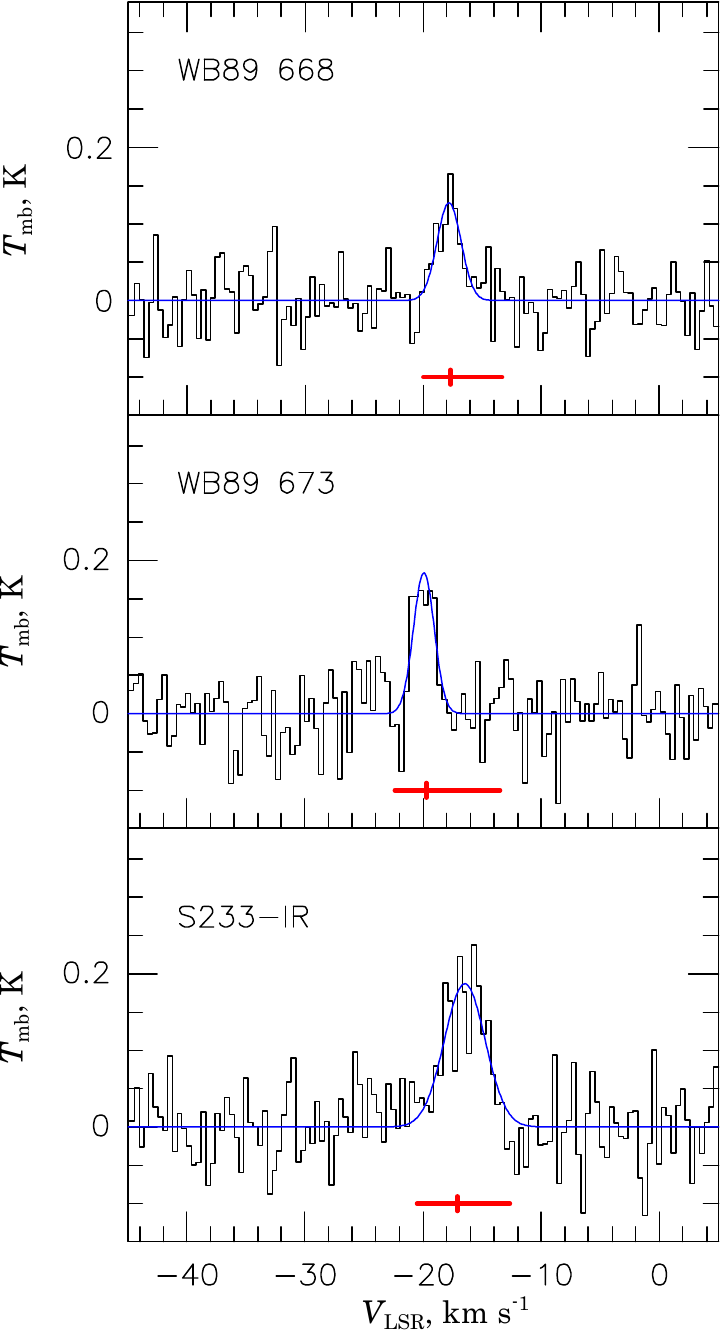}
  \end{tabular}
  \caption[Source spectra in the HC$_3$N line.]{Spectra of the sources detected in the HC$_3$N line at 36.4 GHz.
 The same legend as for Fig.~\ref{img:ch3oh}.}
  \label{img:hc3n}
\end{figure*}

All the observed data in the present work have been obtained at the radio telescope RT-22 of the Puschino Radio Astronomy Observatory (PRAO ASC LPI). We conducted several observational sessions in 2013, 2014, and 2015 using scans of 4--7 min. Data from each scan were calibrated to antenna temperature by the reference signal from the 	noise generator with a known antenna temperature and were corrected for atmospheric absorption. Next we obtained spectra averaged over days, then averaged between days. The antenna temperature $T_{\rm a}$ was reduced to the main beam brightness temperature $T_{\rm mb}$ using the main beam efficiency $\eta_{\rm mb}$. Figures~\ref{img:ch3oh}, \ref{img:hc3n}, and \ref{img:nh3} show the spectra of the obtained lines on $T_{\rm mb}$ scale, and Tables~\ref{tbl:ch3oh}, \ref{tbl:hc3n}, and \ref{tbl:nh3_param} present their observed parameters. To estimate the quality of the  obtained data, we observed the source Dr21 and compared the observations with those presented in~\cite{Tolmachev81,Berulis92,Wilson90}. Calibration accuracy from the result of comparison is about 10--30\%.

\subsection{Observations at the 8~mm wavelength}\label{sec:obs8}
To execute the program of observations, we used a two-channel radiometer of RT-22 of 8 mm wavelength. It is designed for simultaneous observations of two spectral lines in case the line frequencies are in the range of 34 to 38~GHz and the difference between the line frequencies is not greater than 2~GHz. Within the present work, the observations were carried out in the methanol line  (4$_{-1}$--3$_{0}~{\rm E}$, 36.1~GHz) and cyanoacetylene line ($J=$~4--3, 36.3~GHz) using the  spectrum analyser with a constant bandwidth of 50~MHz. Spectral resolution was  24.41~kHz which corresponds to 0.20~km\,s$^{-1}$ for the methanol line frequency. Rest frequency for the CH$_3$OH line  was set 36169.29~MHz and for the HC$_3$N line---36392.33~MHz. We used the ON-ON method of observations based on diagram modulation~\cite{Berulis83}, with which we obtain a double signal as a result. Separation between beams (horns) was 23\arcmin. The half-power beam width (HPBW) was 2\arcmin, the beam efficiency $\eta_{\rm mb}= 0.32$. The system temperature $T_{\rm sys}$ at the time of observations was within the range of 200 to 240~K. With average integration time for each source from 2 to 3\,hours per every day of observations, the total integration time was from 5 to 8\,hours for sources with the detected line and from 1 to 2\,hours for sources without the detected line. The achieved noise level $\sigma_{T_{\rm mb}}$ is in the range of 0.05 to 0.28~K on brightness temperature scale. We checked the observed radial velocity using the source Dr21 (\mbox{$\alpha_{\rm J2000} = 20^{\rm h} 38^{\rm m} 55^{\rm s}$} \mbox{$\delta_{\rm J2000}=42\degr 19\arcmin 23\arcsec$}) which has the methanol line at 36.1~GHz detected earlier in the paper by Liechti et al.~\cite{Liechti96} and the cyanoacetylene line---in the the paper by Tolmachev et al.~\cite{Tolmachev81}. The results of this checking showed that the derived radial velocities of the methanol line (\mbox{$V_{\rm CH_3OH}=-2.73\pm0.01$}\,km s$^{-1}$) and cyanoacetylene line (\mbox{$V_{\rm HC_3N}=-3.08\pm0.06$}\, km s$^{-1}$) in the Dr21 source are within the error margin in accordance with~\cite{Liechti96,Tolmachev81}.

The program {\tt CLASS} from the package {\tt GILDAS}~\cite{Maret11} were used for data reduction. The shape of the line profiles in  methanol and cyanoacetylene spectra was approximated with the standard {\tt GAUSS} method. Owing to a complex structure of the methanol line, two Gaussian functions were used in approximation. 

\begin{table*}
 \setcaptionmargin{0mm} \onelinecaptionsfalse \captionstyle{normal}
\caption{Parameters of the methanol lines (CH$_3$OH) at 36.2~GHz. An asterisk denotes the sources in which the line was detected for the first time. In each detected source two components of emission are presented, the narrow and wide (see the discussion on this in Section~\ref{sec:outflow}). The positions of the associated IRAS sources were chosen as the coordinates for the sources, $t$---integration time, and $1\sigma$---achieved noise level. Estimation errors are given in brackets.} \label{tbl:ch3oh}
\medskip
% \footnotesize
\begin{tabular}{l|c|c|c|c|c|cc} \hline
~~~~~Source        & IRAS  &  $T_{\rm mb}$, & $V$, & $\Delta V$, & $t$, & 1$\sigma $,  \\
& &  K & km\,s$^{-1}$ & km\,s$^{-1}$ & min. & K \\ \hline
WB89\,690 & 05380+3608  & $<$0.08        &  &    & 346  & 0.08    \\
WB89\,668 & 05335+3609  & $<$0.10                       &  &  &  173 & 0.10 \\
WB89\,673* & 05345+3556 & 0.21 (0.04)    & $-$18.51 (0.05)        & 0.3 (0.10) & 500 & 0.05    \\
                &                               & 0.17 (0.04) & $-$18.91 (0.16)   & 3.2 (0.38) & 500  &  0.05  \\
S233-IR & 05358+3543            & 0.47 (0.07)    & $-$15.81 (0.09)                & 1.6 (0.2)~~\, & 386 & 0.08  \\
                &                               & 0.23 (0.07)   & $-$17.7~~ (0.48)          & 3.5 (0.8)~~\, &  386  &  0.08   \\
G173.57+2.43 & 05361+3539       & $<$ 0.28      &  &  &  180 & 0.28   \\
S235-AB         & 05375+3540            & 0.57 (0.04)    & $-$16.34 (0.03)        & 0.42 (0.07)  & 353 &  0.07  \\
                &                               & 0.27 (0.04)    & $-$16.64 (0.15)                & 2.38 (0.34) & 353 &  0.07 \\
S235\,Central E & 05377+3548    & $<$0.15  &  & &   126 & 0.15  \\
S235\,East1 & 05382+3547        & $<$0.13  &  & &  106 & 0.13 \\
S235\,East2 & 05379+3550        & $<$0.13  &  &  &  113 & 0.13 \\
Dr21    & -     & 1.82 (0.06)   & $-$2.73 (0.02)  & 0.49 (0.04) & 333
&  0.09  \\         \hline
  \end{tabular}
%   \normalsize
\end{table*}

\subsection{Observations at the 13~mm wavelength}\label{sec:obs13}

Observations in the ammonia line (NH$_3$\,(1,1) and (2,2), 23.6~GHz) were carried out on the single-channel radiometer of 13.5 mm wavelength. For the transition (1,1), we used a rest frequency of 23694.495~MHz and for the line (2,2)---23722.633~ MHz. We used the ON-ON method based on diagram modulation~\cite{Berulis83}, with a double signal as a result. Separation between beams (horns) was 10\arcmin. The half-power beamwidth (HPBW) was 2\farcm6\ and the beam efficiency factor $\eta_{\rm mb}= 0.38$. Two sessions of observations were conducted, in 2013 and 2015. System temperature during the observations was in the range of 110 to 190~K on the antenna temperature scale.

In 2013, only the NH$_3$\,(1,1) line was observed for initial detection of ammonia emssion in the selected sources. We used the band of the spectrum analyzer with a width of 12.5~Mhz and 2048~channels. Spectral resolution was about 6.1~kHz which corresponds to approximately 0.08~km\,s$^{-1}$  for the rest frequency of the ammonia line. Integration time for the sources was 1--2 hours, the achieved noise level $\sigma_{T_{\rm mb}}$ was in the range of 0.1 to 0.2~K for different sources on the brightness temperature scale.

In 2015, we simultaneously observed two low-level ammonia lines in noise transitions (1,1) and (2,2) for estimating the physical parameters of gas. We used an analyzer band of 50~MHz, thus the spectral resolution was about 24.4~kHz corresponding to approximately  0.31~km\,s$^{-1}$ for the rest frequency of the ammonia line. Central frequency of the band of the spectrum analyzer  was set in the middle between transitions  (1,1) and (2,2) so that both transitions fall within the analyzer band. The distance between frequencies of two transitions of the NH$_3$ line was 28.138~MHz which was enough for their simultaneous detection with the spectrum analyzer with a band width of 50~MHz. Integration time for the sources with the detected line was 9--15 hours and for the sources without the detected line---2--5 hours. The achieved level $\sigma_{T_{\rm mb}}$ for the sources with detection was from 0.01 to 0.04~K.

We approximated the spectra of the ammonia lines using the {\tt NH3(1,1)} method from the {\tt CLASS} package~\cite{Maret11}, and the {\tt SMOOTH} procedure from the same package was used for further smoothing.

\section{RESULTS}\label{sec:res}

\subsection{Masses of Clumps and Amount of Molecular Hydrogen in them}\label{sec:mass}

Determining column density and gas mass in the studied molecular clumps, at large we followed the method described in the paper by Roman-Duval et al.~\cite{Roman-Duval2010} with some changes. Formulas for determination of physical parameters of the clumps are given in Annex~\ref{app:co}. The estimation was made in the approximation of local thermodynamic equilibrium (LTE). To estimate the excitation temperature, we have used the data on emssion in the $^{12}$CO molecule line, as it is optically thick which is proved by the low ratio of intensities of lines \mbox{$I(^{12}$CO$)/I(^{13}$CO$)\approx3$--$6$} as compared to the ratio of molecule abundances  \mbox{$^{12}$CO/$^{13}$CO $\approx50$--$70$}. To estimate the column density of the H$_2$ molecules, we used the emssion in the $^{13}$CO line which is optically thinner than the $^{12}$CO line.

To estimate the mass, we used a number of constants. The ratio of abundances ${\rm CO}/{\rm H}_2=8\times10^{-5}$ is in accordance to Simon ~\cite{Simon01}. The distance to all the clumps is assumed equal to \mbox{$2.1\pm0.5$~kpc} which is the average of the estimated distances to young star cluster from the G174+2.5 GMC from the paper by Camargo et al.~\cite{Camargo11}. For the present distance, the average galactocentric radius is $10.1$~kpc, or $1.26 D_{\odot}$ using the estimation of distance from the Sun to the Galactic center \mbox{$D_{\odot}=8.0$~kpc} from the paper by Reid et al.~\cite{Reid93}. With reference to the paper by Langer and Penzias~\cite{Langer90}, the ratio of abundances  $^{12}$C/$^{13}$C at such galactocentric distance is about $70$, thus, the ratio of abundances \mbox{$^{13}$CO/H$_2 =$\,[CO/H$_2$]/[$^{12}$C/$^{13}$C]$~\simeq 1.14\times10^{-6}$}. This value was used to calculate the H$_2$ column density and clump masses.

%Table 3
\begin{table*}%[!p]
 \setcaptionmargin{0mm} \onelinecaptionsfalse \captionstyle{normal}
\caption{Parameters of the cyanoacetylene lines (HC$_3$N) at 36.4~GHz. An asterisk denote the sources in which the line was detected for the first time. The positions of the associated IRAS sources were chosen as the coordinates for the sources, $t$---integration time, and $1\sigma$---achieved noise level. Estimation errors are given in brackets.}
\label{tbl:hc3n}
\medskip
% \footnotesize
\begin{tabular}{l|c|c|c|c|c|c} \hline
~~~~Source & IRAS & $T_{\rm mb}$, & $V$, & $\Delta V$, & $t$, & 1$\sigma $,  \\
& &  K & km\,s$^{-1}$ & km\,s$^{-1}$ & min. & K \\ \hline
WB89\,690       & 05380+3608 & $<$ 0.09         & & & 346 & 0.09  \\
WB89\,668*      & 05335+3609 & 0.12 (0.03)      & $-$17.80 (0.20) & 2.4 (0.5)~~\, & 460 & 0.06  \\
WB89\,673*      & 05345+3556 & 0.18 (0.04)      & $-$19.94 (0.14) & 2.0 (0.26) & 293 & 0.07 \\
S233-IR         & 05358+3543 & 0.19 (0.07)      & $-$16.4~~ (0.24) & 4.1 (0.5)~~\,  & 353 & 0.05  \\
G173.57+2.43    & 05361+3539 & $<$ 0.19         & & & 180 & 0.19  \\
S235-AB         & 05375+3540 &  $<$ 0.08        & & & 266 &  0.08  \\
S235\,Central E & 05377+3548 & $<$ 0.14         & & & 126 & 0.14  \\
S235\,East1     & 05382+3547 & $<$ 0.12         & & & 106 & 0.12  \\
S235\,East2     & 05379+3550 & $<$ 0.17         & & & 113 & 0.17  \\
Dr21            & -                             &  0.56 (0.06) &
~~$-$3.08 (0.06)& 2.9 (0.16) & 343 &  0.06  \\     \hline
\end{tabular}
% \normalsize
\end{table*}

We used the GaussClump algorithm~\cite{Stutzki1990} to detect the molecular clumps more accurately in the $^{13}$CO line. This algorithm works on the principle of fitting of three-dimensional Gaussians into the observed data cube ``position--position--radial velocity'' (PPV) starting from global emssion maximum. As a threshold for minimum spatial size of clumps ($\theta_{\rm FWHM}$), we set a value of 1\farcm7, because at lower threshold, big molecular clumps are divided into individual components. However, even with such a threshold, some clumps (particularly,  S235 Central and G173.17+2.55) divide into separate components which will be discussed further. Minimum line-width threshold ($\Delta V$) for clumps is  0.8~km\,s$^{-1}$ which is determined by the requirement of extraction of only massive clumps with relatively high dispersion of radial velocity. As a result of using the algorithm, we selected 12~clumps corresponding to the IRAS sources by the criterion described in Section~\ref{sec:selection}. Two clumps divide into individual components; they are S235 Central and G173.17+2.55, on which we will write in the discussion of results. The sizes of clumps are determined as full width at half maximum level ($\theta_{\rm FWHM}$). The mass of clumps is estimated after integrating the clump emssion by the area corresponding to the size of clumps at half-intensity level. Table~\ref{tbl:sources} shows the results of estimation of clump sizes, column density of H$_2$, and mass. Figure~\ref{img:CO} presents the map of distribution of column density of  H$_2$  with the overlaid sizes of molecular clumps.

Virial parameter of clumps  $\alpha_{\rm vir} \equiv M_{\rm vir}/M_{\rm CO}$ and radius $R$ were calculated according to the definition in the paper by Kauffmann et al.~\cite{Kauffmann13}, the formulas for calculation are given in Annex~\ref{app:co}. The distance to all the clumps was assumed equal to $2.1$~kpc and the velocity dispersion was calculated from the $^{13}$CO line width with the formula $\sigma_v=2.35\Delta V_{13}$. Table~\ref{tbl:sources} presents the derived $\alpha_{\rm vir}$ and $R$ values. It should be noted that determination methods for the clump radius $R$ and angular size $\theta_{\rm FWHM}$ presented in Table~\ref{tbl:sources} differ: the first is determined from the clump area with the formula $R=\sqrt{A/\pi}$, the latter---from the width (FWHM) of the Gaussian function fitted in the clump intensity profile.

We determined that average column density of the $^{13}$CO molecule toward the considered molecular clumps of the S231--S235 regions are within the range of \mbox{$1.6\times10^{16}$~cm$^{-2}$} in WB\,690 to
\mbox{$4.8\times10^{16}$~cm$^{-2}$} in S235\,East\,1. Using the above relation\mbox{$^{13}$CO/H$_2 = 1.14\times10^{-6}$}, one can obtain average column density of molecular hydrogen toward the clumps, from \mbox{$1.45\times10^{22}$~cm$^{-2}$} in WB\,690 to
\mbox{$4.21\times10^{22}$~cm$^{-2}$} in S235\,East\,1. The derived values should be regarded as a lower estimate of column density of molecular hydrogen, as the analysis of the CO line emssion within the LTE tends to underestimate actual column densities by a factor of $1.3$ to $7$ in accordance to the paper by Padoan~\cite{Padoan00}. The mass of clumps from the CO data is within the range of $733\,{\rm M}_{\odot}$ in WB89\,690 to $2112\,{\rm M}_{\odot}$ in WB89\,673. The virial parameter $\alpha_{\rm vir}$ varies from $0.33$ in S235\,East\,2 to $1.31$ in \mbox{S233-IR}. Its average value is \mbox{$\overline{\alpha_{\rm
vir}}=0.82$} which at large is indicative of correspondence of virial mass and mass from the $^{13}$CO data.

The excitation temperature derived from the $^{12}$CO data should be interpreted as the temperature of gas near the molecular cloud surface, as the $^{12}$CO line is optically thick. According to the derived data (see Table~\ref{tbl:sources}), the molecular clump WB89\,668 ($14.1$~K) has the ``coldest'' surface and the clump S235\,Central E ($35.6$~K)---the ``warmest'' one. The average surface  temperature of the molecular clumps is $24$~K.

%\subsection{�����-�������� ���⭮�� ��������୮�� ����}\label{sec:dense}
\subsection{Lines-Tracers of Dense Molecular Gas}\label{sec:dense}

In the 36.2~GHz observations, the methanol lines were detected toward the molecular clumps  WB89\,673, S233-IR, and \mbox{S235-AB}. In the clump WB89\,673, the methanol emssion was detected for the first time. In most cases, the line profile shape differs from the Gaussian; and toward S233-IR, the line has an asymmetric structure with a pronounced blue wing. Two emssion components, narrow and wide, can be distinguished in the detected methanol lines. The width of narrow components is within the range of 0.3 to 1.6~km\,s$^{-1}$, for wide components---from 2.4 to
3.5~km\,s$^{-1}$. The positions of the narrow and wide components are shifted relative to each other by a value of 0.3 to 1.8~km\,s$^{-1}$. The difference between radial velocities of methanol and $^{13}$CO lines does not exceed~1.2~km\,s$^{-1}$.

In the cyanoacetylene molecular line, we detected the emssion toward the molecular clumps WB89\,668, WB89\,673, S233-IR, and S235\,Central. The line HC$_3$N in the clumps WB89\,668 and WB89\,673 was detected for the first time. The line profile shapes are close to Gaussian, the average line width is about 2.4~km\,s$^{-1}$, except for S233-IR,  the line width of which is 4.0$\pm$0.5~km\,s$^{-1}$. The difference between radial velocities of the HC$_3$N and CO lines does not exceed~0.5~km\,s$^{-1}$ in the clumps WB89\,668 and WB89\,673. In the clump S233-IR, the HC$_3$N radial velocities are redshifted as compared to the $^{13}$CO radial velocity. The difference between radial velocities is~1.4~km\,s$^{-1}$. 

The ammonia emssion was detected toward the molecular clumps WB89\,668, WB89\,673, G173.57+2.43, S233-IR, S235\,Central, S235\,East1, S235\,East2, and S235-AB,  for  WB89\,668 and WB89\,673---for the first time. Hyperfine  structure of the ammonia lines is detected with an accuracy needed to determine physical conditions in the molecular gas. The difference  between radial velocities of the NH$_3$\,(1,1) and $^{13}$CO lines  does not exceed~0.5~km\,s$^{-1}$ in the clumps WB89\,668, WB89\,673, G173.57+2.43, S235-AB, and S233-IR.

Tables~\ref{tbl:ch3oh}--\ref{tbl:nh3_param} show the parameters of the detected molecular lines.

\begin{figure*}%[p]
  \setcaptionmargin{5mm} \onelinecaptionstrue \captionstyle{normal}
  \center
  \begin{tabular}{@{}cccc@{}}
        \includegraphics [scale=0.8] {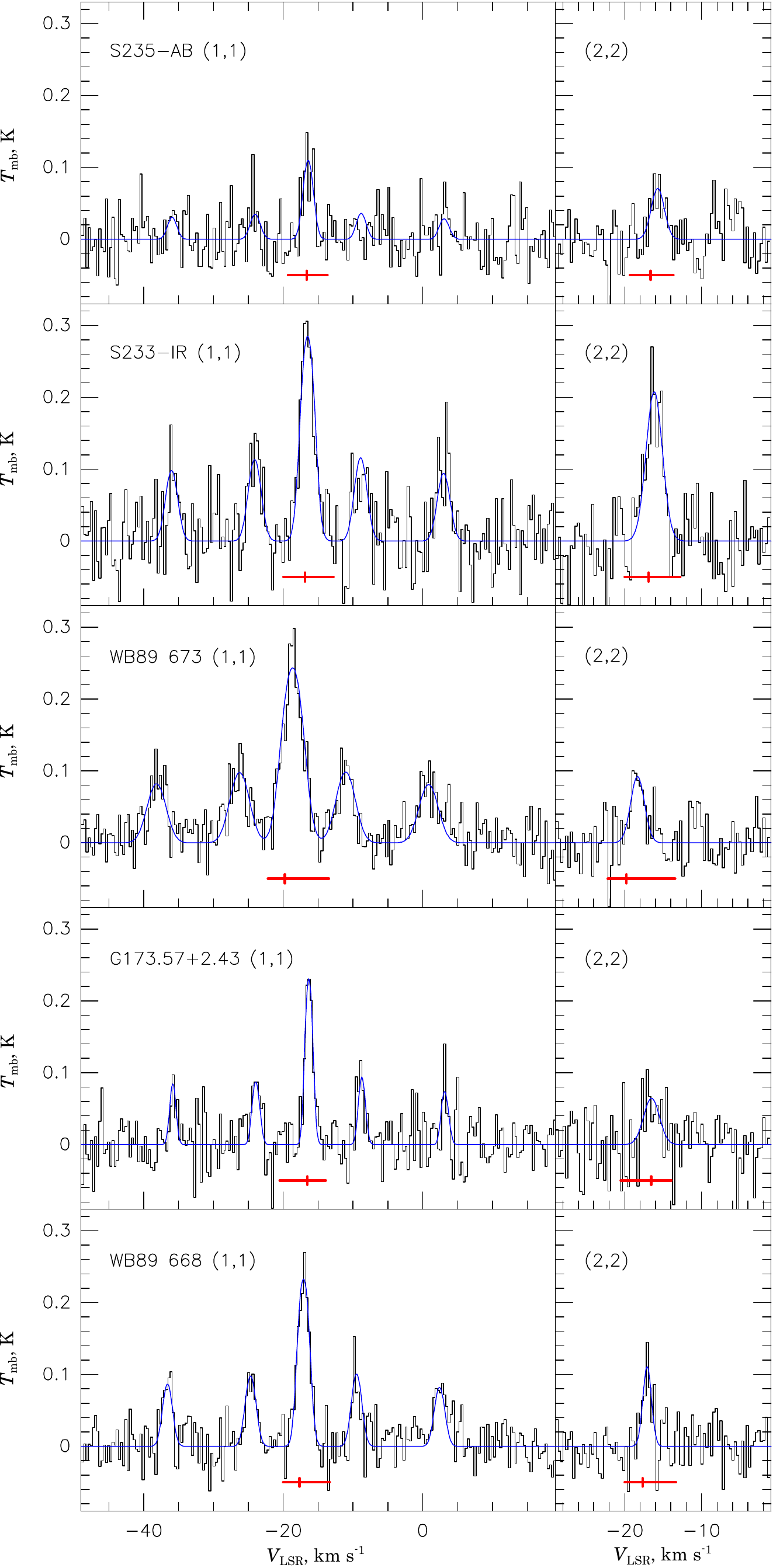} \\
  \end{tabular}
  \caption[Source spectra in the NH$_3$ line.]{Spectra of the detected sourcs in the NH$_3$ line at 23.7~GHz. The same legend as for Fig.~\ref{img:ch3oh}.}
\label{img:nh3}
\end{figure*}

%Table 4
\begin{table*}%[p]
 \setcaptionmargin{5mm} \onelinecaptionsfalse \captionstyle{normal}
\caption{Results of observations of star-forming regions in the ammonia line  (NH$_3$) at a frequency of 23.7~GHz and physical parameters of gas in the molecular clumps (estimation errors are in brackets). An asterisk marks the sources, in which the line was detected for the first time.  The positions of the $^{13}$CO emission peak were chosen as the coordinates for the sources (see the coordinates from Table~1.}
\label{tbl:nh3_param}
\medskip
% \footnotesize
\begin{tabular}{l|c|c|c|c|c|c|c|c|c|c} \hline
~~~~Source & $(J,K)$ & $T_{\rm mb}$, & $T_{\rm B}$, & $V$,        & $\Delta V$,  &$\tau_{\rm(1,1)m}$& \multirow{2}{*}{$\frac{\Theta_{\rm beam}^{2}}{\Theta_{\rm maj}\Theta_{\rm min}}$} & $T_{\rm kin}$, & $N({\rm NH_3}),$      & $n{\rm (H_{2})}$, \\
             &         & K             & K            & km\,s$^{-1}$& km\,s$^{-1}$ &                  &         &  K           & $10^{14}$~cm$^{-2}$  &  $10^3$~cm$^{-3}$ \\
%\hline
\qquad (1)                &(2)    &(3)          &(4)        &(5)          &(6)            &(7)         &(8)  &(9)         &(10)        &(11) \\
\hline
WB89\,668*         & (1,1) & 0.24 (0.01) & 1.3 (0.1) & $-$17.1 (0.1) & 1.7 (0.1)     & 1.4 (0.3)  & 4.3 & 16.5 (1.9)~~ & ~~7.3 (1.5)  & 4.2 \\
                   & (2,2) & 0.09 (0.01) & 0.5 (0.1) & $-$17.1 (0.1) & 1.5 (0.3)     &            &     &            &            &     \\
WB89\,673*         & (1,1) & 0.25 (0.01) & 1.1 (0.1) & $-$18.6 (0.1) & 3.1 (0.1)     & 1.2 (0.3)  & 3.5 & 15.9 (1.5)~~ & 12.4 (2.2) & 4.1 \\
                   & (2,2) & 0.08 (0.01) & 0.4 (0.1) & $-$18.2 (0.2) & 2.9 (0.6)     &            &     &            &            &     \\
S233-IR            & (1,1) & 0.28 (0.03) & 2.2 (0.2) & $-$16.5 (0.1) & 2.0 (0.1)     & 1.2  (0.4) & 6.9 & 29.4 (11.8)& ~~9.9 (1.9)  &7.2  \\
                   & (2,2) & 0.21 (0.04) & 1.7 (0.3) & $-$16.2 (0.1) & 2.6 (0.3)     &            &  &  & &    \\
G173.57+2.43       & (1,1) & 0.23 (0.03) & 1.4 (0.2) & $-$16.3 (0.1) & 1.2 (0.1)     & 1.2 (0.5)  & 5.1 & 14.4 (5.2)~~ & ~~5.1 (1.0)  & 5.6 \\
                   & (2,2) & 0.06 (0.01) & 0.4 (0.1) & $-$16.5 (0.3) & 2.3 (0.5)     &            &  &  & &    \\
S235-AB            & (1,1) & 0.11 (0.03) & 0.6 (0.2) & $-$16.4 (0.1) & 1.6 (0.3)     & 0.6  (0.9) & 4.4 & 27.4 (59.7)& ~~2.2 (1.7)  & 2.8 \\
                   & (2,2) & 0.07 (0.04) & 0.4 (0.21 & $-$15.7 (0.3) & 2.1 (0.6)     &            &  &  & &    \\

\hline
  \end{tabular}
%   \normalsize
\end{table*}

\subsection{Temperature and Density of Molecular Gas}\label{sec:cond}

To estimate the temperature and density of gas on the assumption of the local thermodynamic equilibrium (LTE), we used the ratio of the antenna temperature of the NH$_3$ spectrum main component to the antenna temperature of the satellite components of the hyperfine  structure of the spectrum and the ratio between the main components of the  transitions \mbox{NH$_3(J,H)=$\,(1,1)} and~(2,2). The method of determination of physical parameters is described in Annex~\ref{app:nh3}. As a result, we derived estimations of the ammonia column density ($N_{{\rm NH_3}}$), kinetic temperature ($T_{\rm kin}$), and number density of molecular gas ($n_{\rm H_2})$.

For the reason that the source size in the NH$_3$ line in the present complex (about 50--110\arcsec, see Table~6 from Kirsanova et al.~\cite{Kirsanova14}) is smaller than the RT-22 beam size (156\arcsec), then it is necessary to apply a correction for filling of the beam. We used the following formula (see equations 8.21--8.22 from Rohlfs and Wilson~\cite{Rohlfs04}) to estimate the brightness temperature of a source:

\begin{equation}
% \begin{array}{l}
T_{\rm B}=T_{\rm mb} \times \left( 1+\frac{\theta_{\rm
beam}^2}{\theta_{\rm maj}\theta_{\rm min}}  \right)
%  \end{array}
 \label{eq:Tb}
\end{equation}
\noindent where $T_{\rm mb}=T_{\rm a}/\eta_{\rm mb}$---the main-beam brightness temperature, $\theta_{\rm beam}=152''$---the size of the RT-22 beam at a wavelength of 1.35~mm, $\theta_{\rm maj}$ and $\theta_{\rm min}$---the sizes of sources (FWHM) along major and minor axes.

In order to estimate sizes of sources in the NH$_3$ line, we used the data on emssion in the continuum at a wavelength of 1.1~mm from the Bolocam survey \cite{Ginsburg13}. The comparison of emssion sources in the continuum at a wavelength of 1.1~mm and in the NH$_3$ line toward S235\,Central, East\,1, East\,2~\cite{Kirsanova14}, \mbox{S233-IR}~\cite{Zinchenko97}, and G173.57+2.43~\cite{Schreyer96} showed that their sizes correspond to each other, thus, for brightness temperature estimation in the NH$_3$ line, one can use source sizes at a wavelength of 1.1~mm. The source sizes were determined with fitting of two-dimensional Gaussians into images by the {\tt IMFIT} method from the {\tt MIRIAD} package~\cite{Sault95}. The determination results are presented in Table~\ref{tabl:sizes}. For calculating the physical parameters of gas from the NH$_3$ lines, we used brightness temperatures obtained from formula~(\ref{eq:Tb}).

%Table 5
\begin{table}
 \setcaptionmargin{0mm} \onelinecaptionsfalse \captionstyle{normal}
\caption[Clump sizes from the data in the continuum at a wavelength of 1.1~mm.]{Clump sizes toward the star-forming regions S231--S235  from the data in the continuum at a wavelength of 1.1~mm from the Bolocam survey~\cite{Ginsburg13}. Root-mean-square deviation is given in brackets.} \label{tabl:sizes}
\medskip
% \footnotesize
%\begin{tabular}{l|c|c|c|c|c|c|c|c|c|c} \hline
\begin{tabular}{l|c|c} \hline
\multicolumn{1}{c|}{Source}      & $\theta_{\rm maj}$, & $\theta_{\rm min}$,   \\
                        & $\arcsec$  & $\arcsec$  \\
\hline
WB89\,668    & 87 (12) & 65 (10) \\
WB89\,673    & 149 (13) & 66 (5)  \\
G173.57+2.43 & 81 (11) & 59 (8)  \\
S233-IR      &67 (2)  & 53 (2)  \\
S235\,Central& 163 (11)& 96 (7)  \\
S235\,East1  & 125 (12)& 55 (4)  \\
S235\,East2  & 93 (14) & 86 (13) \\
S235-AB      & 86 (3)  & 65 (10) \\
\hline
\end{tabular}
\end{table}

Table~\ref{tbl:nh3_param} shows the derived physical parameters. Table legend: $T_{\rm mb}$---main-beam brightness temperature, $T_{\rm B}$---brightness temperature of a source, $\tau_{\rm(1,1)m}$---optical depth of the main component of the (1,1) line. Column~8 gives a coefficient to convert from $T_{\rm mb}$ into $T_{\rm B}$ using formula~(\ref{eq:Tb}). Number density of molecular gas toward the clumps  WB89\,673, WB89\,668, S233-IR, G173.57+2.43, and S235-AB is within the range of $2.8$--$7.2\times 10^3$~cm$^{-3}$. The highest gas number density ($n=7.2\times10^3$~cm$^{-3}$) was detected toward the  S233-IR. The kinetic temperature for the clumps  WB89\,668, WB89\,673 and G173.57+2.43 is nearly similar and equal to $14$--$16$~K, and for S233-IR and S235-AB---$27$--$30$~K. The column density of ammonia $N_{\rm NH_3}$ is within the range of $2.2$--$12.4\times10^{14}$~cm$^{-2}$.

The brightest emission of ammonia was detected in S233-IR and WB89\,673, and the gas temperature in these clumps is different: \mbox{$T_{\rm kin}=29.4\pm11.8~$~K} and \mbox{$T_{\rm kin}=15.9\pm1.5~$~K}. It is notable that symmetric components of the hyperfine components in S233-IR have different intensities which is indicative of the effects of deviation from the LTE that we will discuss futher.

\section{DISCUSSION}\label{sec:disc}

Detection of all molecular clumps in the GMC G174+2.5 allows us to study conditions and sequence of star formation in it. The morphology of star-forming regions is complex, gas distribution is inhomogeneous which can be seen from Fig.~\ref{img:CO}. In the paper by Heyer et al.~\cite{Heyer96}, they have concluded that there are molecular filaments connected with the regions  S235 and S231. In the paper by Evans et al.~\cite{Evans81}, they say that there are two molecular clouds with different radial velocities around the H\,II S235  region. In the paper by Kirsanova et al.~\cite{Kirsanova08,Kirsanova14}, there is an interpretation of the kinematic structure of the S235 neighborhood  within the model of triggered star formation scenario ``collect-and-collapse'' (see~\cite{Elmegreen77,Whitworth94a}). In the paper by Ladeyshchikov et al.~\cite{Ladeyschikov15}, it is shown that in the S233  region, expansion of the  H\,II region has caused a collapse of a massive molecular clump initiating further star formation in it.

\subsection{Dense Gas Distribution in the Regions S231--S235}

\begin{table}%[!p]
 \setcaptionmargin{0mm} \onelinecaptionsfalse \captionstyle{normal}
\caption[Detection of lines in sources.]{List of initial detection of molecular lines from the literature data and the results of this paper (marked with [tw]). Symbols $+$ and $-$ denoted the sources with and without the detected line respectively}
\label{tbl:registrations}
\medskip
% \footnotesize
\begin{tabular}{l|c|c|c|c|c} \hline
Source                &H$_2$O   & NH$_3$ & HC$_3$N & CH$_3$OH & SiO  \\
                        & 22 GHz  & 23 GHz & 36 GHz & 36 GHz & 86 GHz   \\
\hline
WB89\,690       & +~\cite{Wouterloot93}         &               &               &   + [tw]   & $-$~\cite{Harju98} \\
WB89\,668       &  +~\cite{Wouterloot93}        & + [tw]         & + [tw]         &               &  \\
WB89\,673       &  +~\cite{Wouterloot93}        & + [tw]         & + [tw]         & + [tw]         & +~\cite{Harju98} \\
G173.17+2.55    &               & $-$ [tw]       &               &               & \\
G173.57+2.43    & +~\cite{Wouterloot93}         & +~\cite{Wouterloot88}         &               &               &  \\
S233-IR                 &  +~\cite{Wouterloot93}        & +~\cite{Schreyer96}   & +~\cite{Pirogov03}    & +~\cite{Liechti96}    & +~\cite{Harju98} \\
S235\,Central   &  +~\cite{Wouterloot93}        & +~\cite{Schreyer96}   & +~\cite{Pirogov03}    &               & $-$~\cite{Harju98} \\
S235\,East1     & $-$ [tw]      & +~\cite{Kirsanova14}          &               &               &         \\
S235\,East2     & $-$ [tw]  & +~\cite{Kirsanova14}       &               &               & +~\cite{Harju98} \\
S235-AB                 &  +~\cite{Wouterloot93}        & +~\cite{MacDonald81}  & +~\cite{Pirogov03}    & +~\cite{Liechti96}    & $-$~\cite{Harju98} \\
\hline
\end{tabular}
\end{table}

In this paper, we show that spatial distribution of emssion in the CO molecular lines can be used to study general characteristics of molecular clouds and to search for dense clumps with possible star formation in them. However, one single CO molecule is insufficient for the detailed study of star formation, as the emssion in the CO molecule is indicative of the presence of  medium-density gas and becomes saturated with higher density. In the case of a extended region of gas through the line of sight, its column density can be high from the CO-line estimations, however, the gas is not of high density actually. One of the most striking examples of such a situation can be observed in NGC\,6334~\cite{Walsh10}, where the peak in the CO line does not correspond to the peaks in the HCO$^+$, HCN and N$_2$H$^+$ molecular lines. To prove the presence of high-density gas, additional observations in lines of molecules with high critical density are required. In the present paper, we use the lines of the HC$_3$N~(4--3) and NH$_3$~(1,1) molecules, the critical densities of which are  $n_{\rm
crit}\simeq10^4$~cm$^{-3}$ and $n_{\rm crit}\simeq10^3$~cm$^{-3}$ respectively. Table~\ref{tbl:registrations} shows the  literature data on the detection of lines-tracers of dense gas in the regions S231--S235. The  HC$_3$N abundance is significant in  dense heated gas near young stars (see, e.g., Meier et al.~\cite{Meier05} and  Lindberg et al.~\cite{Lindberg11}), moreover, the HC$_3$N lines in star-forming regions are optically thin, as it was shown by Van~den~Bout~\cite{VandenBout83}.

The star-forming regions S231--S235 in the cyanoacetylene line  were studied earlier in the paper by Alakoz et al.~\cite{Alakoz02}, but the emssion at a level of $3\sigma=0.25$~K was not detected for the sources \mbox{S233-IR} and \mbox{S235-AB}. In a year, however, the lines for three sources were detected in the paper by Pirogov et al.~\cite{Pirogov03}. These sources were designated by the authors as S231, S235B, and S235 and associated with the molecular clumps \mbox{S233-IR}, \mbox{S235-AB}, and S235\,Central respectively. 

In accordance with the paper by Meier et al.~\cite{Meier05}, the cyanoacetylene emssion is in good agreement with the emssion in the continuum at 3~mm. According to the BGPS archive data~\cite{Ginsburg13}, all the molecular clumps with the detected emssion in the HC$_3$N line also radiate in the continuum at 1.2~mm. Nevertheless, we have not detected any emssion toward S235\,East1, S235\,East2, and S235\,Central, which can be associated  with insufficient detection threshold of the observations.

For the first time, the ammonia line in S231--S235 was detected toward the source S235-AB almost at the same time by  Ho et al.~\cite{Ho81} and MacDonald et al.~\cite{MacDonald81}. Later, Harju et al.~\cite{Harju91} detected the ammonia line in the  IRAS 05361+3539 source (our G173.57+2.43), Schreyer et al.~\cite{Schreyer96}---in S233-IR, S235\,Central, S235-AB, and Zinchenko et al.~\cite{Zinchenko97}---in the source S233-IR. Then in the paper by Kirsanova et al.~\cite{Kirsanova14}, the region of S235 was studied in detail, also they obtained the maps of ammonia radio-brightness distribution and determined physical parameters of gas toward the following clusters: S235\,East1, S235\,East2, S235\,Central, and S235-AB. 

As seen from Fig.~\ref{img:nh3}, the relation between the brightnesses of hyperfine components in S233-IR  is irregular, i.e., it differs from the relation in the LTE conditions. In the paper by Stutzki et al.~\cite{Stutzki85}, it is shown that such an anomaly occurs when several small gas clumps fall into the beam of a telescope.

Virial stability in molecular clumps is described in detail in the paper by Kauffmann et al.~\cite{Kauffmann13}. It is usually supposed that if the virial parameter  \mbox{$\alpha_{\rm vir} > \alpha_{\rm cr}$}, then a clump or a molecular cloud  is gravitationally stable. If \mbox{$\alpha_{\rm vir} \lesssim \alpha_{\rm cr}$}, then density and pressure perturbation of a clump can cause a gravitational contraction of matter and starting star-formation processes. For isothermal clumps with the Jeans mass without considering magnetic fields---\mbox{$\alpha_{\rm cr}\simeq2$}~\cite{Kauffmann13,McKee92}. As seen from Table~\ref{tbl:sources}, all the considered clumps  \mbox{$\alpha_{\rm vir} \lesssim 2$}, which is indicative of their gravitational instability.

Some molecular clumps in the CO line have a complex extended structure that is not exactly described by the GaussClump algorithm, which supposes the Gaussian brightness distribution. The analysis shows that some molecular clumps divide into multiple components even with the use of a quite high threshold (1\farcm7) for minimum source size. 	First of all, the clump G173.17+2.55 is referred to such clumps; it is a filament by structure and, thus, divides into the individual clumps NE and SW. Analogously, the molecular clump S235-Central is divided into two separate components (S235-Central E and S235-Central W), which agrees with the NH$_3$ emssion structure from the paper by Kirsanova et al.~\cite{Kirsanova14}. To study the spatial-kinematic structure of such clumps, the observations of better resolution are required as well as applying of other methods of structure detection, such as GetFilaments~\cite{Menshchikov13},
FIVe~\cite{Hacar13}, and DisPerSe~\cite{Sousbie11}, etc.

\subsection{Star Formation in Molecular Clumps} \label{sec:star.formation}

According to the modeling data from the paper by Clark et al.~\cite{Clark14},  the average column density of molecular clumps should exceed $10^{21}$~cm$^{-2}$ to let star-formation processes start in them. Our paper shows that average gas column density in clumps from the data on the  $^{13}$CO emssion is within the range of $1.4$ to $4.3\times10^{22}$~cm$^{-2}$, thus, all the clumps under study  are candidates   star-forming regions.

Young star clusters in the GMC G174+2.5 were studied in the paper by Camargo et al.~\cite{Camargo11} using the photometry with the 2MASS data\footnote{Two-Micron All-Sky Survey, available here {\tt www.ipac.caltech.edu/2mass/releases/allsky/}}. They reported on 14 young star clusters embedded in molecular gas. Young star clusters toward the molecular clumps WB89\,673 and WB89\,668 have not been detected in their paper. All the clusters found are indicated with dashed circles in Fig.~\ref{img:CO}. 

In accordance with~\cite{Camargo11}, age, position, and sizes of young star clusters near  S235 (S235\,Central, East1, and East2) are in agreement with the scenario of the ``collect-and-collapse'' star formation process. Color characteristics of stars toward the molecular clumps S235-AB, S232-IR, \mbox{S233-IR}, and G173.57+2.43 match the characteristics of the embedded young star clusters. It is supposed that the age of these clusters is 3--5 Myr (see Table~2 from~\cite{Camargo11}) and they have not ``scattered'' the 	surrounding molecular gas yet.

Clusters in the S235-AB region were studied in the paper by Felli et al.~\cite{Felli97,Felli04,Felli06}. They showed that the star cluster is located between the nebulae S235A and S235B, the first of which is an H\,II{} region. It is shown in the paper by Boley et al.~\cite{Boley09} that S235~B is a reflection nebula. They found a layer of heated molecular gas from the southern part of S235A. This gas is located between the H\,II{}  region and the molecular cloud. In the more recent paper, Felli et al.~\cite{Felli06} show that there are young stars in this layer. They deduced that the interaction of S235A and the surrounding molecular cloud probably caused the formation of the second generation of stars in this region. 

Thus, young star clusters was detected toward all the considered molecular clumps except for G173.17+2.55, which are: S235\,Central, S235\,East1, S235\,East2, S235-AB, S232-IR,
S233-IR, G173.57+2.43, WB89\,673, and WB89\,668. The presence of  young star clusters is indicative of active star-forming processes in the considered molecular clumps.

\subsection{Evidence of Outflows in Molecular Clumps} \label{sec:outflow}

In this Section, we will discuss the evidence of outflows in the studied massive clumps based on characteristics of maser emssion of molecules.

Maser emssion of methanol is a distinctive feature of star-forming regions. In the early studies of Batrla~\cite{Batrla87} and Menten~\cite{Menten91}, two types of methanol masers were distinguished empirically. Type~II~masers (e.g., at 6.7, 12, 37.7, and 107~GHz) pumped by IR emission of dust from  young stars (see the papers by Sobolev et al.~\cite{Sobolev05,Cragg05}), thus, they are detected in the vicinity of such objects. Type~I~masers (e.g., at 36, 44, and 95~GHz) emerge as a result of collisional radiative pumping (see Sobolev et al.~\cite{Sobolev07}) and usually indicate the presence of gas compressed by a shock wave. Such gas is often detected near young stars with an outflow interacting with ambient matter (see Voronkov et al.~\cite{Voronkov06}). Emergence of molecular outflows from star-formation regions is an essential stage of star-formation process~\cite{Bachiller96}. Type~I~masers are normally detected at some distance from young stellar objects, as is seen in the papers of  Kurtz et al.~\cite{Kurtz04}, Voronkov et al.~\mbox{\cite{Voronkov06,Voronkov14}}. However, type~I methanol masers can arise in any region of the interstellar medium, where moderate-speed shock waves emerge: with the collisions of molecular clouds (see Salii et al.~\cite{Salii02}), with the supernova explosions  (see Pihlstrom et al.~\cite{Pihlstrom14}), in places of interaction of  H\,II{} regions and surrounding molecular gas (see  Voronkov et al.~\cite{Voronkov10b}), and in the  regions with complex hydrodynamic motions  (see  Voronkov et al.~\cite{Voronkov10a}).
 		 
In the methanol line at a frequency of 36.2~GHz, the emssion toward S231--S235 was first detected in the paper by Haschick et al.~\cite{Haschick89}, where they obtained the spectrum of the S235 source corresponding to the molecular clump S235-AB in the present paper. Later, in the paper by Liechti et al.~\cite{Liechti96}, maser and thermal methanol lines were detected in two sources: S233-IR and S235-AB. In the methanol lines which we detected at 36~GHz in the sources  WB89\,673, S233-IR, and S235-AB, two components can be distinguished, wide and narrow. The wide component ($\geq
2.0$~km\,s$^{-1}$) is often interpreted as ``thermal'' and the narrow component  ($\leq 1.5$~km\,s$^{-1}$)---as ``maser''~\cite{Liechti96}. According to the conclusions by Berulis et al.~\cite{Berulis92}, the ``maser'' component can originate both in protostar neighborhood and in a usual gravitationally stable fragment of the interstellar medium emerging due to turbulence. Type~I methanol masers can also be associated with protostars at the early stages of collapse (see Sobolev et al.~\cite{Sobolev83}, Sutton et al.~\cite{Sutton04}) and form at the borders of hypercompact H\,II{} regions (see Sobolev et al.~\cite{Sobolev07}). Thus, a methanol maser can not be a definitive criterion of the presence of outflows from young stellar objects at early star-forming stages.

It should be noted that there is a restriction to methanol formation in shock waves. In accordance with the paper by Garay et al.~\cite{Garay02}, methanol can not exist in shock waves moving with speeds exceeding 10~km\,s$^{-1}$, as these molecules decay at high speeds. As an additional indicator of shock waves associated with outflows from young stellar objects, one can use the SiO molecular lines, see the papers by Schilke et al.~\cite{Schilke97} and Caselli et al.~\cite{Caselli97}. With the shock wave propagation, the methanol abundance of the gas phase increases owing to evaporation from the dust-particle surface~\cite{Salii02}. Compared to methanol, the SiO molecule does not decay at high speeds of shock waves  (from 10 to 40~km\,s$^{-1}$, see~\cite{Schilke97}), thus, emssion in  the SiO lines is a reliable tracer of outflows. The existence of outflows from the young molecular clumps WB89\,673 and S233-IR is proved by the presence of the SiO outflow in the paper by Harju et al.~\cite{Harju98}. The outflow in S233-IR has been studied earlier in the literature (see \mbox{\cite{Porras00,Beuther02,Khanzadyan04,Ginsburg09}}), where they show that it emerges in the massive-star forming region. The SiO emssion in S235-AB was not detected in~\cite{Harju98}, although, there is the emssion observed in the maser methanol lines in this clump at 36~GHz~\cite{Haschick89,Liechti96}.

In most cases, the maser emssion of water is detected in star-forming regions (see Section 6.1.1 from Gray~\cite{Gray12}). The presence of water masers toward the molecular clumps in the S231--S235 regions is additional evidence of  active star-formation processes ongoing in them. The H$_2$O masers were detected toward the clumps WB89\,690, WB89\,668, WB89\,673, G173.57+2.43, S233-IR, S235\,Central, and S235-AB (see Table~\ref{tbl:registrations}). 

The outflow in the S235-AB region was  proved in a series of papers by Felli et al.~\cite{Felli97,Felli04,Felli06}. It is shown in the paper by Shepherd and Watson~\cite{Shepherd02} that at least two young stellar objects from the cluster are responsible for matter outflow in the G173.58+2.43 region.

Thus, molecular outflow evidence are found in the following massive clumps: WB89\,690, WB89\,668, WB89\,673, G173.57+2.43, S233-IR, S235\,Central, S235\,East2, and S235-AB.
They are not found toward S235\,East1 and G173.17+2.55.

\section{CONCLUSION}\label{sec:concl}

\begin{list}{}{
\setlength\leftmargin{2mm} \setlength\topsep{2mm}
\setlength\parsep{0mm} \setlength\itemsep{2mm} }
\item (1) Based on the archive data on the CO molecule, we identified 10 massive star-forming clumps in the giant molecular cloud G174+2.5. All of them belong to the star-forming regions  S231--S235 which took their names from the H\,II regions located in them.

\item (2) The clumps are gravitationally unstable and, from the CO data, their masses are within the range of about~700 to 2000~$M_{\odot}$.

\item (3) As a result of observations, we obtained the spectra of the methanol, cyanoacetylene, and ammonia lines: 

\begin{itemize}
\item The emssion in the cyanoacetylene line at 36.4~HGz was detected toward three molecular clumps: WB89\,668, WB89\,673, and S233-IR. It indicates high density of molecular gas. The HC$_3$N column density appear to be sufficient for being detected on the RT-22 radio telescope.

\item The methanol line at 36.2~HGz was detected toward WB89\,673, S233-IR, and S235-AB. Detection of emssion in this line denotes the shock fronts in the clumps.

\item The emssion in the ammonia line was detected toward 6 clumps: WB89\,668, WB89\,673, S233-IR, G173.57+2.43, S235\,AB, and S235\,Central. Physical parameters of the molecular clumps were determined from the ammonia line: temperature, column density of ammonia, and molecular gas number density. It was determined that the clump temperature is within the range of 16 to 30~K and the molecular gas number density ---of 2800 to 7200~cm$^{-3}$.

\end{itemize}
% \medskip
% \medskip

\item (4) Embedded young stellar clusters are found in all the clumps except for G173.17+2.55.

\item (5) Molecular outflow evidence are observed toward all massive clumps except for S235\,East1 and G173.17+2.55. 
\end{list}
% \end{enumerate}

\begin{acknowledgments}

The study was supported with the program 211 of the Government of the Russian Federation, agreement \No 02.A03.21.0006. The work was partially supported by the Ministry of Education and Science of the Russian Federation (state assignment \No 3.1781.2014/K). 

D.~A.~Ladeyshchikov is thankful to G.~T.~Smirnov, V.~A.~Gusev, and S.~V.~Logvinenko for guidance, support, and help in development of the automation system for the RT-22 double-channel radiometer. 

M.~S.~Kirsanova is grateful to the Russian Science Foundation (grant MK-2570.2014.2)  and the PSD-15 program of the Physical Sciences Department of the Russian Academy of Sciences.

A.~P.~Tsivilev thanks the PSD program ``Interstellar and Intergalactic Medium: Active and Extended Objects''.

\end{acknowledgments}

\bibliography{biblio}{}
\bibliographystyle{AstroBull}

\appendix

\section{A. ESTIMATION OF MOLECULAR CLOUD MASSES FROM THE CO RADIO LINES}\label{app:co}

The presented method is intended to determine physical parameters of molecular clouds from the data in the CO lines. In general, the method corresponds to the one in the paper by Roman-Duval et al.~\cite{Roman-Duval2010} with some changes concerning the detection of column densities and clump masses. To estimate the column density, we used the LTE assumption. It was also supposed that the $^{12}$CO lines are optically thick. In this case, the excitation temperature $T_\rmn{ex}$ of the  line can be determined from the radiative transfer equation (formula 15.29 from~\cite{Rohlfs04}) for the \mbox{$^{12}$CO(1--0)} line:

\begin{equation}
\begin{aligned}
T_{\rm ex} = 5.53 / \ln \left(1+\frac{5.53}{T_{\rm B}^{12}+0.837}
\right)
\end{aligned}
\label{A1}
\end{equation}
\noindent where $T_{\rm B}^{12}$---the brightness temperature of the  \mbox{$^{12}$CO(1--0)} line. In the present case, the microwave background $T_{\rm bg}=2.7~$K is taken into account. As the average size of sources in the $^{12}$CO and $^{13}$CO (from~1\farcm9 to~4\farcm1, see Table~\ref{tbl:sources}) is bigger than the beam width (45\arcsec),  \mbox{$T_{\rm B}=T_{\rm mb}$}. 

To determine the optical depth and CO column density, we used the isotope of the $^{13}$CO molecule, as it is less abundant, and its optical depth is smaller as compared to $^{12}$CO. The latter results in the fact that emssion is less subject to saturation effects. The optical depth of the $^{13}$CO(1--0) line can be derived with formula (15.31) from~\cite{Rohlfs04}:

\begin{equation}
\begin{aligned}
\tau_0^{13} = & - \ln \left[ 1- \frac{T_{\rm B}^{13}}{5.3} \left\{  \left[ \exp \left(\frac{5.3}{T_{\rm ex}} \right)-1 \right]^{-1}  \right. \right. \\
& \Biggl. \Biggl. - 0.16  \Biggr \}^{-1} \Biggr ],
\end{aligned}
\label{A2}
\end{equation}
\noindent where $T_{\mathrm{B}}^{13}$---the brightness temperature of the \mbox{$^{13}$CO(1-0)} line, $T_{\rm ex}$---the excitation temperature. For the CO linear molecule, the excitation is characterized by the same temperature $T_{\rm ex}$, thus, the column density $N$ and optical depth $\tau$ are connected with the following relation (see equation~3 from~\cite{Roman-Duval2010}):

\begin{equation}
\begin{aligned}
N(^{13}{\rm CO})=2.6&\times10^{14} \frac{T_{\rm
ex}}{1-\exp(-5.3/T_{\rm ex})} \\ & \times \int \tau^{13}(v){\rm d}v,
\end{aligned}
\label{A3}
\end{equation}
\noindent 
moreover, in the case of the Gaussian profile of optical depth $\int
\tau^{13}(v){\rm d}v=\tau^{13}_0\sigma_v\sqrt{2\pi}$, where 
$\sigma_v$ is the $^{13}$CO line velocity dispersion, $\tau_0$ is the optical depth in the center of the line. The H$_2$ column density is derived from the abundances of$^{12}$CO/$^{13}$CO and also CO/H$_2$:
\begin{equation}
\begin{aligned}
N({\rm H}_2)=N(^{13}{\rm CO})\times {\rm \frac{^{12}CO}{^{13}CO}}
\left[{\rm \frac{CO}{H_2}} \right]^{-1}
 \label{A4}
\end{aligned}
\end{equation}
The relation CO/H$_2$ $\simeq 8\times10^{-5}$ is in accordance with~\cite{Blake97}. The relation $^{12}$CO/$^{13}$CO varies from 40 to 70 depending on the distance from the Galactic center to the source according to ~\cite{Langer90}. At a distance of 2~kpc from the Sun in the direction to the galactic anticenter, the ratio of abundances is $^{12}$CO/$^{13}$CO $\simeq$~70.

The mass is determined as  a integration of column density distribution $N$(H$_2$)  on the source surface ${\rm d}A$:
\begin{equation}
\begin{array}{rcl}
 M=\mu m_{\rm H_2}\int{N_{\rm H_2}{\rm d}A}= \mu m_{\rm H_2}D^2\int{N_{\rm H_2}{\rm d}\Omega},
 \end{array}
 \label{A5}
\end{equation}
where $\mu$ is the ratio of the interstellar gas to the mass of the hydrogen molecule, \mbox{$\mu\approx1.33$}~\cite{Hildebrand83},
$m_{\rm H_2}$ is the mass of the hydrogen molecule and the surface element is ${\rm d}A$ related with the solid angle through the relation \mbox{${\rm d}A=D^2{\rm d}\Omega$}, where $D$ is the distance to the source in kpc. 

When combining expressions~(\ref{A4}) and~(\ref{A5}) and insertion of numerical constants, we obtain the following expression for the source mass:
\begin{equation}
\begin{aligned}
{M}  = 0.41 D^2 \int_{\alpha,\delta}{\frac{T_{\rm
ex}\tau^{13}\sigma_v\sqrt{2\pi}}{1-\exp(-5.3/T_{\rm ex})}}
\Delta\alpha\Delta\delta~~({M_{\rm \odot}})
\end{aligned}
 \label{A6}
\end{equation}

\noindent where $T_{\rm ex}=T_{\rm ex}(\alpha,\delta)$ is the excitation temperature in the given map cell which is calculated with formula (\ref{A1}), $\tau^{13}=\tau^{13}(\alpha,\delta)$ is the optical depth of $^{13}$CO in the center of the line profile of the given map cell which is calculated with  formula~(\ref{A2}), $\sigma_v=\sigma_v(\alpha,\delta)$ is the $^{13}$CO line-profile velocity dispersion, $\Delta \alpha$ and $\Delta \delta$ is the size of a map cell along $\alpha$ and $\delta$ expressed in arcminutes. A coefficient preceding the integral, 0.41, differs from the value 0.27 accepted in the paper by Roman-Duval et al.~\cite{Roman-Duval2010} due to using a different ratio of abundances $^{12}$CO/$^{13}$CO (70 instead of 45).

The virial parameter of the clumps $\alpha_{\rm vir} \equiv M_{\rm vir}/M$ is calculated in accordance with the definition given in~\cite{Kauffmann13}:
\begin{equation}
\begin{aligned}
\alpha_{\rm vir}   =  \frac{5\sigma^2_v R}{GM} = & 1.2 \left(
\frac{\sigma_v}{{\rm km\,s^{-1}}} \right)^2   \left(\frac{R}{{\rm
pc}} \right) \left( \frac{M}{{\rm M_{\odot}}} \right)^{-1}
 \end{aligned}
 \label{A7}
\end{equation}

The clump radii are derived as in~\cite{Roman-Duval2010} from the area which is occupied by a clump at the full width of half maximum:
\begin{equation}
\begin{aligned}
 R =\sqrt{\frac{A}{\pi}}=\sqrt{\frac{\Omega D^2}{\pi}}=\sqrt{ \frac{N_{\rm pix}\Delta\alpha\Delta\delta D^2}{\pi}},
 \end{aligned}
 \label{A8}
\end{equation}
\noindent where $N_{\rm pix}$ is the number of map cells occupied by a clump, $\Delta\alpha$ and $\Delta\delta$ are the map cell sizes. The velocity dispersion $\sigma_v$ in the $^{13}$CO molecular line is calculated with the formula according to~\cite{Roman-Duval2010}:
\begin{equation}
\begin{aligned}
\sigma^2_v = \frac{\sum{T_{13}(v-\bar{v})^2}}{\sum{T_{13}}},
 \end{aligned}
 \label{A9}
\end{equation}
\noindent and addition in formula (A9) is done along $(\alpha,\delta,v)$ for those $T_{13}$ values that exceed the intensity level $4\sigma$.

\section{B. METHOD OF ESTIMATION OF MOLECULAR GAS PHYSICAL PARAMETERS FROM THE AMMONIA RADIO LINES}\label{app:nh3}

The emssion spectrum of the ammonia line is described in detail in~\cite{Ho_83}. Here we present the end formulae only which we have used when estimating the gas temperature, its density, and ammonia abundance in it. It is known that populations of metastable levels (1,1) and (2,2) are determined by collisions which allows us to use the assumption on the LTE and estimate the kinetic gas temperature  $T_{\rm gas}$. Using the {\tt NH3(1,1)} method from the  {\tt CLASS} package, one can estimate the optical depth of the main component $\tau_{\rm (1,1) m}$. Further, the excitation temperature of transition (1,1) is determined which is supposed to be similar for all other inversion transitions in the LTE conditions:
\begin{equation}
\begin{array}{rcl}
T_{\rm ex} = \frac{T_{\rm B (1,1)}}{1-\exp(-\tau_{\rm (1,1) m})} +
T_{\rm bg}~~({\rm K}),
\end{array}
\end{equation}
\noindent Table~\ref{tbl:nh3_param} shows the brightness temperature of the main component of (1,1) line, $T_{\rm B (1,1)}$. The $T_{\rm bg}$ value is the background temperature which is determined by the temperature of the cosmological microwave background and the background emssion of a source $T_{\rm s}$:
\begin{equation}
\begin{array}{rcl}
T_{\rm bg}=2.73 + T_{\rm s} ~~({\rm K}).
\end{array}
\end{equation}
The column density of ammonia at level (1,1) is estimated as in~\cite{Mangum92}:
\begin{equation}
\begin{aligned}
 N_{\rm (1,1)} = 6.60 \times & 10^{14} \frac{T_{\rm ex}}{\nu_{(1,1)}} \tau_{\rm (1,1) m}  \\ \times & \Delta V_{\rm (1,1)}~~({\rm cm^{-2}}),
\end{aligned}
\end{equation}
\noindent where $\Delta V_{\rm (1,1)}$ is the (1,1) line width in km\,s$^{-1}$, its value is given in Table~\ref{tbl:nh3_param}, $\nu_{(1,1)} = 23.7$ is the transition (1,1) frequency in GHz.

The total column density of ammonia in the LTE conditions in terms of  four low levels: the ground one and (1,1), (2,2), and (3,3) is equal to:
\begin{equation}
\begin{aligned}
N_{\rm NH_3}=N_{(1,1)} & \left(  \frac{1}{3} e^{\frac{21.3}{T_{\rm
rot}}} + 1 + \frac{5}{3} e^{\frac{-41.2}{T_{\rm rot}}} \right. \\  &
\left.  + \frac{14}{3} e^{\frac{-99.4}{T_{\rm rot}} } \right)~~({\rm
cm^{-2}}),
\end{aligned}
\end{equation}
\noindent where $T_{\rm rot}$ is the rotational temperature determining the relation of populations between levels~(2,2) and~(1,1).

In the LTE conditions~\cite{Ho_83}:
\begin{equation}
\begin{aligned}
T_{\rm rot}  = & -41.5  {\rm ln} \left( \frac{-0.282}{\tau_{\rm (1,1)
m}} {\rm ln} \left(1 - \frac{T_{\rm B (2,2) m}}{T_{\rm B (1,1) m}}
\right. \right. \\ & \biggl. \biggl. \times  \left(1-e^{-\tau_{\rm
(1,1) m}}\right) \biggr) \biggr)^{-1}~~({\rm K}).
\end{aligned}
\end{equation}

To estimate $T_{\rm gas}$, one uses the relation obtained in~\cite{Walmsley_1983} with the coefficients from~\cite{Danby_1988}:
\begin{equation}
\begin{aligned}
T_{\rm rot} = \frac{T_{\rm gas}}  {1+ \frac{T_{\rm gas}}{41.5}\ln
\left( 1+ \frac{C_{(2,2\rightarrow 2,1)}}{C_{(2,2\rightarrow 1,1)}}
\right)}~~({\rm K}),
\end{aligned}
\end{equation}
\noindent The coefficients $\rm C_{(2,2\rightarrow 2,1)}$ and $\rm C_{(2,2\rightarrow 1,1)}$ are given in a tabulated form in~\cite{Danby_1988} for several $T_{\rm gas}$ values. Equation $({\rm B}6)$ can be solved with the iteration method.

If we know $T_{\rm ex}$, we can estimate the molecular gas number density $n{\rm (H_{2})}$ taking into consideration only two levels of inversion transition (1,1) with the use of the following relation~\cite{Ho_83}:
\begin{equation}
\begin{array}{rcl}
n{\rm (H_{2})} = & \frac{\rm A_{\rm (1,1)}}{\rm C_{\rm (1,1)}} \left(
\frac{J_{\nu}(T_{\rm ex})-J_{\nu}(T_{\rm bg})}{J_{\nu}(T_{\rm
gas})-J_{\nu}(T_{\rm ex})} \right) \\  & \times \left(
1+\frac{J_{\nu}(T_{\rm gas})}{{\rm h} {\rm \nu}_{\rm (1,1)} / {\rm
k}} \right)~~(\rm cm^{-3}),
\end{array}
\end{equation}
\noindent where
\begin{equation}
\begin{aligned}
J_{\nu}(T) = \frac{{\rm h} \nu_{\rm (1,1)}}{\rm k} \left( \exp
\frac{{\rm h} \nu_{\rm (1,1)}}{{\rm k} T} - 1\right).
\end{aligned}
\end{equation}
\noindent  The coefficient $\rm A_{\rm (1,1)} = 1.7 \times
10^{-7}$~s$^{-1}$~\cite{Rohlfs04}, and $\rm C_{\rm
(1,1)}$ is taken from Table in~\cite{Danby_1988} for the certain $T_{\rm gas}$ value.

\end{document}